# Field-induced Polar Order at the Néel Temperature of Chromium in Rare-earth Orthochromites: Interplay of Rare-earth and Cr Magnetism

B. Rajeswaran[1], D. I. Khomskii[2], A. K. Zvezdin[3], C. N. R. Rao[1] and A. Sundaresan[1*]

[1]*Chemistry and Physics of Materials Unit and International Centre for Materials Science, Jawaharlal Nehru Centre for Advanced Scientific Research, Jakkur P.O., Bangalore 560 064, India*
[2]*II. Physikalisches Institut, Universität zu Köln, ZülpicherStrasse 77, 50937 Köln, Germany*
[3]*Prokhorov General Physics Institute, Russian Academy of Sciences, ul. Vavilova 38, Moscow, 119991 Russia*
[*]*sundaresan@jncasr.ac.in*

## Abstract

We report field-induced switchable polarization (P ~ 0.2 – 0.8 μC/cm$^2$) below the Néel temperature of chromium ($T_N^{Cr}$) in weakly ferromagnetic rare-earth orthochromites, $RCrO_3$ (R=rare-earth) but only when the rare-earth ion is magnetic. Intriguingly, the polarization in $ErCrO_3$ ($T_C$ = 133 K) disappears at a spin-reorientation (Morin) transition ($T_{SR}$ ~ 22 K) below which the weak ferromagnetism associated with the Cr-sublattice also disappears, demonstrating the crucial role of weak ferromagnetism in inducing the polar order. Further, the polarization (P) is strongly influenced by applied magnetic field, indicating a strong magnetoelectric effect. We suggest that the polar order occurs in $RCrO_3$, due to the combined effect of poling field that breaks the symmetry and the exchange field on R-ion from Cr-sublattice stabilizes the polar state. We propose that a similar mechanism could work in the isostructural rare-earth orthoferrites, $RFeO_3$ as well.

Magnetoelectric multiferroics constitute an emerging class of novel materials that combine coupled electric and magnetic dipole orders [1-6]. Interaction between the two order parameters leads to magnetoelectric effect, which gives rise to magnetization on application of an electric field or to electric polarization on applying a magnetic field. Coupling between magnetic and ferroelectric order is generally strong in the so-called type-II materials where the ferroelectricity arises due to certain type of magnetic spin structures [4]. A cycloidal spin structure below the spiral magnetic ordering of $Mn^{3+}$ ions in the centrosymmetric orthorhombic (P*bnm*) $TbMnO_3$, [5,6] and collinear magnetic ordering with E-type magnetic structure in the $HoMnO_3$ [7] break the inversion symmetry causing net electric polarization. It has been reported that the induced moment on the rare-earth ion enhances the ferroelectric polarization in $DyMnO_3$ [8]. In cycloidal-spin induced ferroelectricity, the antisymmetric exchange Dzyaloshinskii-Moriya (DM) interaction ($S_i \times S_j$) plays a key role in producing the electric polarization [9]. Based on symmetry analysis of magnetoelectric interactions in rare-earth orthoferrites and orthochromites, it has been shown that the spontaneous electric polarization or magnetic-field-induced polarization can appear at the magnetic ordering temperature of rare-earth ions [10]. This effect was found experimentally in some rare-earth orthoferrites, $RFeO_3$ (R=Gd and Dy) where exchange striction between R and Fe-moments has been suggested to responsible for the origin of ferroelectricity [11,12,13].

It has been reported recently that $SmFeO_3$, which belongs to the rare-earth orthoferrite family, exhibits ferroelectric features at the magnetic ordering temperature of iron ($T_N^{Fe}$ = 670 K) [14]. This is surprising because rare-earth moments order at low temperatures ( < 10 K) and thus the exchange striction mechanism as discussed above may not directly account for the origin of polarization at such high temperature. Also, the spin-current or the inverse DM interaction model for such a canted antiferromagnetic system suggests zero net polarization because local polarization cancels out due to the alternate arrangement of pairs of canted spins [5,9]. We believe that magnetic interactions between $Sm^{3+}$ and canted moments of $Fe^{3+}$ ions may play a crucial role in inducing ferroelectric behaviour in this oxide. In fact, anisotropic interactions between R and $Fe^{3+}$ ions were suggested to be responsible for the temperature induced spin-reorientation in this, as well as in many other rare-earth orthoferrites ($RFeO_3$) [15,16]. This is consistent with the fact that such spin reorientation does not occur in orthoferrites with nonmagnetic R-ion as in $YFeO_3$. Accordingly, the orthoferrites with a non-magnetic R-ion should not show ferroelectric polarization at the magnetic transition. However, the spin disorder at the Fe-site in $YFe_{1-x}M_xO_3$ (M=Cr, Mn) can induce spin-reorientation [17-19].

Since the high Néel temperature of $RFeO_3$($T_N^{Fe}$ =620-740 K) makes it difficult to carry out polarization measurements due to the high leakage current, we have chosen isostructural rare-earth orthochromites ($RCrO_3$) which exhibit similar magnetic properties with relatively lower $T_N$ values (120-300 K). In this letter, we report field-induced polar order below the Néel temperature of chromium in the weakly ferromagnetic $RCrO_3$ containing magnetic R-ions such as Sm, Gd, Tb, Tm and Er and no polar order when the R-ion is non-magnetic.

In the orthochromites with the orthorhombic structure (P*bnm*), the following three types of G-type antiferromagnetic configurations are observed: $\Gamma_1(A_x, G_y, C_z)$, $\Gamma_2(F_x, C_y, G_z)$ and $\Gamma_4(G_x, A_y, F_z)$ following the Bertaut notation [20]. The first configuration, as reported in $ErCrO_3$ below a spin-reorientation transition ($T_{SR}$ ~ 22 K), does not allow weak ferromagnetism but the second and third have weak ferromagnetism along the x and z-directions, respectively [21-25]. When the R-ion is nonmagnetic, the ground state remains weakly ferromagnetic with the magnetic spin configuration, $\Gamma_4$. If the R-ion is magnetic, the high temperature magnetic structure can be $\Gamma_4$ or $\Gamma_2$ depending upon the R-ion. At low temperatures, some $RCrO_3$ systems show a spin-reorientation transition below which the spin structure becomes $\Gamma_2$ or $\Gamma_1$ [21]. All the three spin configurations ($\Gamma_1$, $\Gamma_2$ and $\Gamma_4$) are shown in Fig. 1.

Polycrystalline samples of $RCrO_3$ (R=Sm, Gd, Tb, Er, Tm, Lu and Y) were prepared by the solid state reaction of stoichiometric quantities of $R_2O_3$ and $Cr_2O_3$ at 1673 K for 12 hours followed by several intermittent grinding and heating. Phase purity was confirmed by Rietveld refinement on the X-ray powder diffraction data collected with Bruker D8 Advance diffractometer. Magnetic measurements were carried out with a vibrating sample magnetometer (VSM) in the physical property measurement system (PPMS) of Quantum Design, USA. Capacitance and pyroelectric measurements were carried out with LCR meter Agilent E4980A and 6517A Keithley electrometric resistance meter, respectively, using multifunction probe in PPMS. In the pyroelectric current measurement, first the sample was poled by applying an electric field of +1.43 kV/cm and -1.43 kV/cm at a temperature (greater than $T_N$) and then the sample was cooled down to low temperatures under the applied field. After shorting the circuit for a reasonably long duration, the current was measured using the electrometer while warming the sample to a temperature higher than the $T_N$ at a rate of 4 K/min. Upon integrating the pyrocurrent with respect to time and dividing it by the area of the sample, we obtain electric polarization, which can be plotted as a function of temperature. Positive-Up and Negative-Down (PUND) and resistivity measurements were carried out by using Radiant Technologies Inc., precision workstation.

In Figure 2(a), we show the temperature dependence of magnetization of $SmCrO_3$ under field-cooled conditions with an applied field of 100 Oe. The observed behavior is characteristic of antiferromagnetic ordering of Cr-moments ($T_N$ = 197 K) with weak ferromagnetism. The drop in magnetization below 40 K is due to spin-reorientation where the Cr-spin configuration changes from $\Gamma_4(G_x, A_y, F_z)$ to $\Gamma_2(F_x, C_y, G_z)$. Spin-reorientation in orthoferrites and orthochromites is known to be brought about by anisotropic magnetic interactions between $R^{3+}$ and Fe/Cr ions [15, 25]. In Figure 2 (b), we show the temperature dependent dielectric constant, $\varepsilon_r(T)$ of $SmCrO_3$ measured at 1 kHz. The dielectric constant is almost independent of temperature up to 170 K above which it increases with a steep raise above 200 K which is associated with a large frequency dependent Maxwell-Wagner relaxation [19]. Though we do not observe a dielectric anomaly in the vicinity of $T_N$, the derivative of $\varepsilon_r$ (T) shows (right inset) a clear anomaly at $T_N$ indicating magnetodielectric effect [26]. On the other hand, a clear anomaly in $\varepsilon_r$ is seen at $T_{SR}$ (left inset). It is seen from the insets that the frequency dependence of $\varepsilon_r$ is large above 200 K but relative small at low temperatures. A similar anomaly in $d\varepsilon_r/dT$ is observed at $T_N$ in Tb, Gd and Tm orthochromites which do not exhibit spin-reorientation down to 15 K [27].

Figure 2 (c) shows electric polarization of $SmCrO_3$ (P ~ ± 0.80 μC/cm$^2$ at T = 15 K for E = ±1.43 kV/cm) and pyroelectric current peaks (inset) obtained from pyroelectric measurements, where it is to be noted that the polarization results are corrected for leakage current. This figure clearly demonstrates the development of the electric polarization in the vicinity of the $T_N^{Cr}$ and confirms that the polarization is switchable. Further, it shows a signature of spin-reorientation in pyroelectric current and electric polarization. It is important to note that the observed polarization values are relatively higher than those reported in most of the magnetically induced ferroelectric materials [2, 26-30]. This figure also demonstrates that the effect of magnetic field on the polarization where an increase of 0.05 μC/cm$^2$ at 2 T is observed at 15 K.

We also find electric polarization near $T_N$ = 167 K for the canted antiferromagnetic $GdCrO_3$ (P ~ ± 0.7 μC/cm$^2$ at T = 15 K for E = ±2.25 kV/cm) as shown in Fig. 3 [31, 32]. In the case of $TbCrO_3$, we observe electric polarization (P ~ ± 0.5 μC/cm$^2$ at T = 15 K for E = ±1.43 kV/cm) in the vicinity of $T_N$ =157 K [27]. In $TmCrO_3$, in addition to electric polarization at $T_N$ = 127 K (P ~ ± 0.25 μC/cm$^2$ at T = 15 K for E = ±1.43 kV/cm), we also observe a temperature-induced magnetization reversal with a compensation temperature ($T^*$) of 25 K indicating that the Tm and canted Cr moments are coupled antiferromagnetically similar to that observed in $SmFeO_3$ [14, 27].

In order to further confirm the intrinsic nature of the observed electric polarization, we have performed PUND measurements [27, 28]. This method allows separating out the intrinsic contribution to pyrocurrent from that due to leakage currents, etc. Though the value of remanent polarization obtained by this method is small (0.01 μC/cm$^2$), probably due to relatively small applied effective poling field, these measurements confirm that the polarization observed in orthochromites with magnetic rare-earth ion below $T_N$ is intrinsic, and that these systems are indeed multiferroic. It is noteworthy that in the case of $LuCrO_3$ and $YCrO_3$ with nonmagnetic R-ion no intrinsic electric polarization is observed [27].

In addition to the magnetic R-ion, we show that the weak ferromagnetism of the Cr-sublattice is also essential for inducing polarization in this orthochromites. The compound, $ErCrO_3$ undergoes antiferromagnetic ordering with a weak ferromagnetism at $T_N^{Cr}$ = 133 K (Fig. 4a) and at low temperatures it exhibits a spin-reorientation (Morin) transition at $T_{SR}$ ~ 22 K below which the weak ferromagnetism disappears (Fig. 1a). We observe dielectric anomalies both at $T_N$ and at $T_{SR}$ confirming the magnetoelectric coupling (Fig. 4b). Interestingly, the electric polarization (Fig. 4c), measured following the procedure adapted for CuO, [33] disappears along with the weak ferromagnetism below $T_{SR}$, which demonstrates that the weak ferromagnetism of Cr-sublattice is essential for inducing electric polarization. However, the polarization below $T_{SR}$ is revived by cooling the sample under an applied magnetic field (± 2 T). The applied magnetic field suppresses the spin-reorientation and thus the weak ferromagnetic state and the polarization remain down to the lowest temperature.

Most importantly, the electric polarization is influenced strongly by applied magnetic field indicating a strong coupling between spontaneous magnetization and polarization as shown in Fig. 5 for $GdCrO_3$. By applying a positive magnetic field, the polarization increases whereas it decreases under negative applied field. It can be seen that the change in polarization is ~ 0.1 $\mu C/cm^2$ per tesla.

Our experimental results clearly establish the presence of electric polarization (P ~ 0.2 – 0.8 $\mu C/cm^2$) below the magnetic ordering temperature of Cr-ions ($T_N^{Cr}$ = 120 – 300 K) in $RCrO_3$ but only when the R-ion is magnetic and the ordered Cr-sublattice is weakly ferromagnetic. It is apparent that these phenomena could be observed in similar materials, for example, the isostructural orthoferrites, $SmFeO_3$ with magnetic R-ion. The experimental data on $SmFeO_3$ agree with our findings, and they show in addition that the polarization in $SmFeO_3$, and most probably also in our systems, lies in the ab-plane [14].

Theoretical interpretation of our findings should, first of all, take into account symmetry restrictions imposed by the crystal structure. As argued e.g. in [ Refs. [10], [34], and especially stressed in [35], with full symmetry (P*bnm*) no magnetic ordering of Cr (or Fe) in $RCrO_3/RFeO_3$ can in itself produce ferroelectric polarization. As demonstrated by our experimental results, the observed polarization should be connected with the interplay of magnetism between Cr and R. Spontaneous magnetization of R below their Néel temperature can in some cases give polarization [10,12]. But at higher temperatures $T_N^R < T < T_N^{Cr}$ only induced magnetization on R-ions due to their interaction with Cr could exist, and, with the full lattice symmetry intact, this induced magnetization also can not lead to ferroelectricity [36].

Thus, we are left with two options: 1) One can suspect that the actual lattice symmetry of orthochromites and orthoferrites is lower than the usually assumed P*bnm* (or P*nmc*), due to some tiny, as yet undetected distortion. In this case the Cr-R exchange field could in principle lead to non-zero polarization. The R ion in the P*bnm* structure are not lying at the centre of symmetry, their point group is Cs not $D_{2h}$: they lie in the mirror plane perpendicular to the c-axis, but their coordinates in this ab-plane are not fixed, and each R ion can be shifted so as to produce a local dipole. However, the P*bnm* symmetry requires that such dipoles at other R ions are opposite, so that the sum of such dipoles in a unit cell is zero. Thus, there is no net ferroelectric polarization caused by orthorhombic distortions, but one can consider such perovskites as antiferroelectric. Consequently, also the contribution to such dipoles e.g. due to exchange striction of R-Cr exchange also sum up to zero. However, if the real symmetry of the lattice (and consequently the point symmetry of the R-ion) would be lower, these contributions would not cancel and could give ferroelectric polarization below $T_N^{Cr}$.

Another option, which we think is more plausible, is that the very process of measuring polarization, which involves the poling procedure, causes some small distortion of R ions and their surrounding, producing odd contribution to the crystal field of R and triggering, "releasing'' the exchange-striction mechanism. That is, we propose that the poling field reduces the symmetry, and the metastable state thus formed can survive after we release the poling field. The main part of polarization arises due to the R-Cr exchange striction, which strongly enhances the distortion initially caused by the poling field. Thus, the poling acts as a trigger, and the metastable "self-poled" state created below $T_N^{Cr}$ is the state which displays a relatively large measured polarization. This picture agrees with the general experimental conclusions summarized above.

Now, the dominant contribution to the exchange striction comes from the isotropic Heisenberg-like R-Cr exchange. A purely antiferromagnetic G-type ordering of Cr sublattice would not lead to a net non-zero striction, but if there exists weak ferromagnetism of Cr, there would be nonzero striction of the same sign at every (magnetic) R ion, i.e. there would appear net polarization lying in ab-plane.

In principle, there may also appear effects due to an antisymmetric R-Cr interaction [15], which could also contribute to polarization. The analysis of these terms should be done specifically for particular system. But, in general, we expect that such contributions to polarization would be weaker, so that the main conclusions rationalized above, generally hold.

In conclusion, we found that the field-induced metastable state with electric polar order appears at the magnetic ordering temperatures of $Cr^{3+}$ ions in the weakly ferromagnetic rare-earth orthochromites ($RCrO_3$, where R is a magnetic rare-earth ion), exhibiting a relatively large electric polarization ~ 0.2 – 0.8 $\mu C/cm^2$, starting at rather high temperatures (~120 – 250 K) corresponding to the Néel temperatures of the Cr subsystem. We propose that the multiferroic behaviour of these systems is caused by the interaction between magnetic rare earth and weak ferromagnetic $Cr^{3+}$-ions following the breaking of symmetry by the effect of poling. The observation of a strong dependence of polarization with applied magnetic field along with their high Curie temperatures and large polarization suggest that these materials are potential candidates for device applications.

The authors are grateful to P. Mandal and Maxim Mostovoy for useful discussion and help. The work of D.Kh. was supported by the German programs SFB 608 and FOR 1346, and by the European project SOPRANO.

**Figure captions**

Fig. 1. (Color online) Spin structures in $\Gamma_1$, $\Gamma_2$ and $\Gamma_4$ with P*bnm* symmetry showing absence of spin canting in $\Gamma_1$. The location of R-ion is marked by spheres without spin.

Fig. 2. (Color online) (a) Field-cooled magnetization of $SmCrO_3$ at 100 Oe with respect to temperature. (b) Dielectric constant with respect to temperature at 1 kHz. Insets show the frequency dependence of the dielectric constant and the first derivative of dielectric constant in the vicinity $T_{SR}$ and $T_N$, respectively. (c) Electric polarization (corrected for leakage) as a function of temperature and the inset shows the corresponding pyroelectric current.

Fig. 3. (Color online) (a) Field-cooled magnetization of $GdCrO_3$at 100 Oe with respect to temperature. (b) Electric polarization (corrected for leakage) as a function of temperature. Inset in (b) shows the pyroelectric current as a function of temperature with (red) and without the presence of magnetic field of 2T (black).

Fig.4(Color online) (a) Field cooled magnetization of $ErCrO_3$ at 100 Oe showing a spin reorientation transition from weakly ferromagnetic $\Gamma_4$ to collinear $\Gamma_1$ magnetic spin structure at ~22 K. (b) Dielectric constant as a function of temperature at 5 kHz. Insets show the dielectric anomaly at both the $T_{SR}$ and $T_N$ temperature regions. (c) Ferroelectric polarization as a function of temperature measured following the procedure adapted for CuO in ref.[33]. It should be noted that the ferroelectricity disappears in the collinear phase.

Fig. 5.(Color online) Effect of applied magnetic field on electric polarization in $GdCrO_3$. It can be seen that the polarization changes systematically with the strength and polarity of the applied magnetic field.

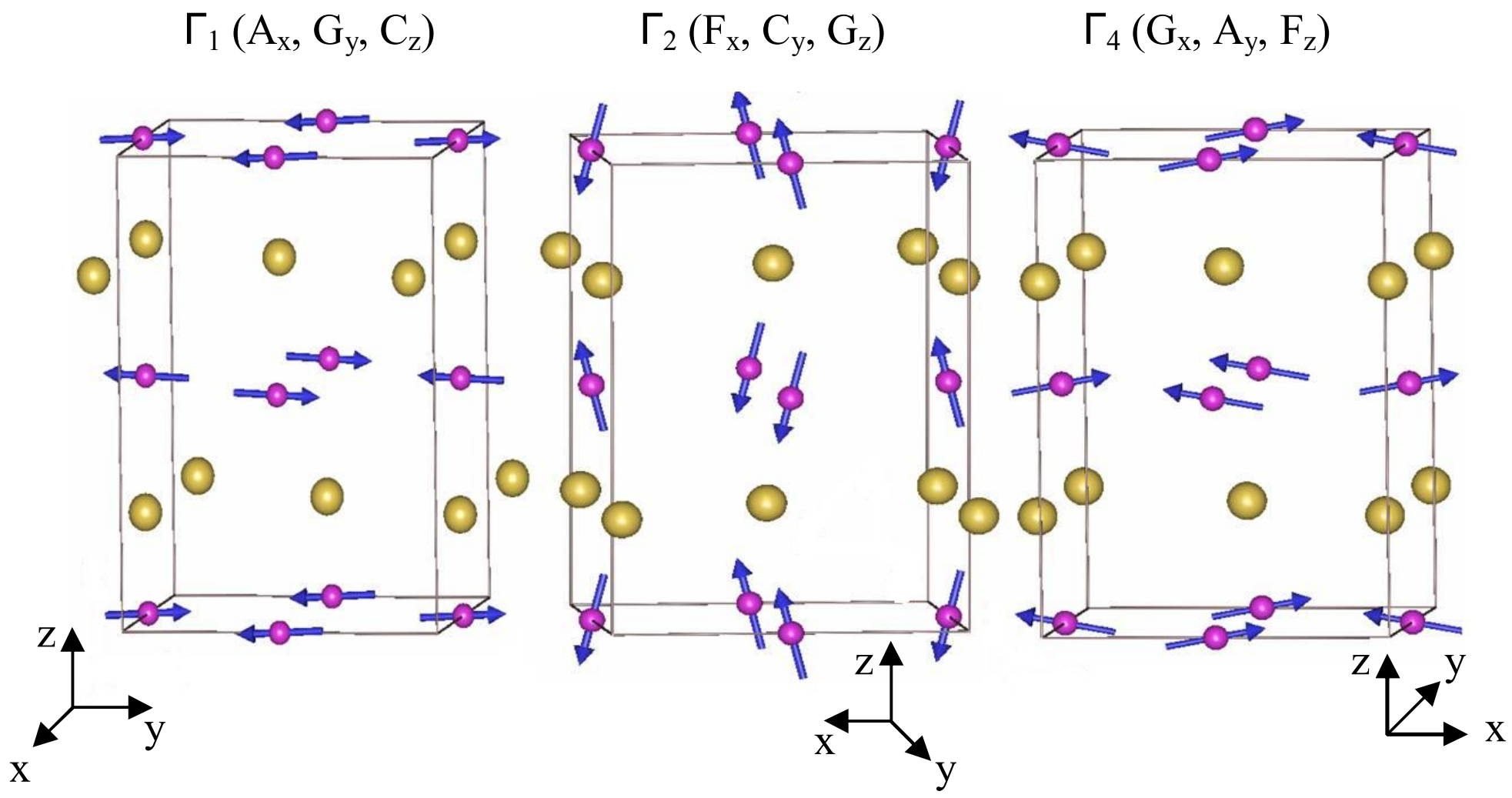

Fig. 1

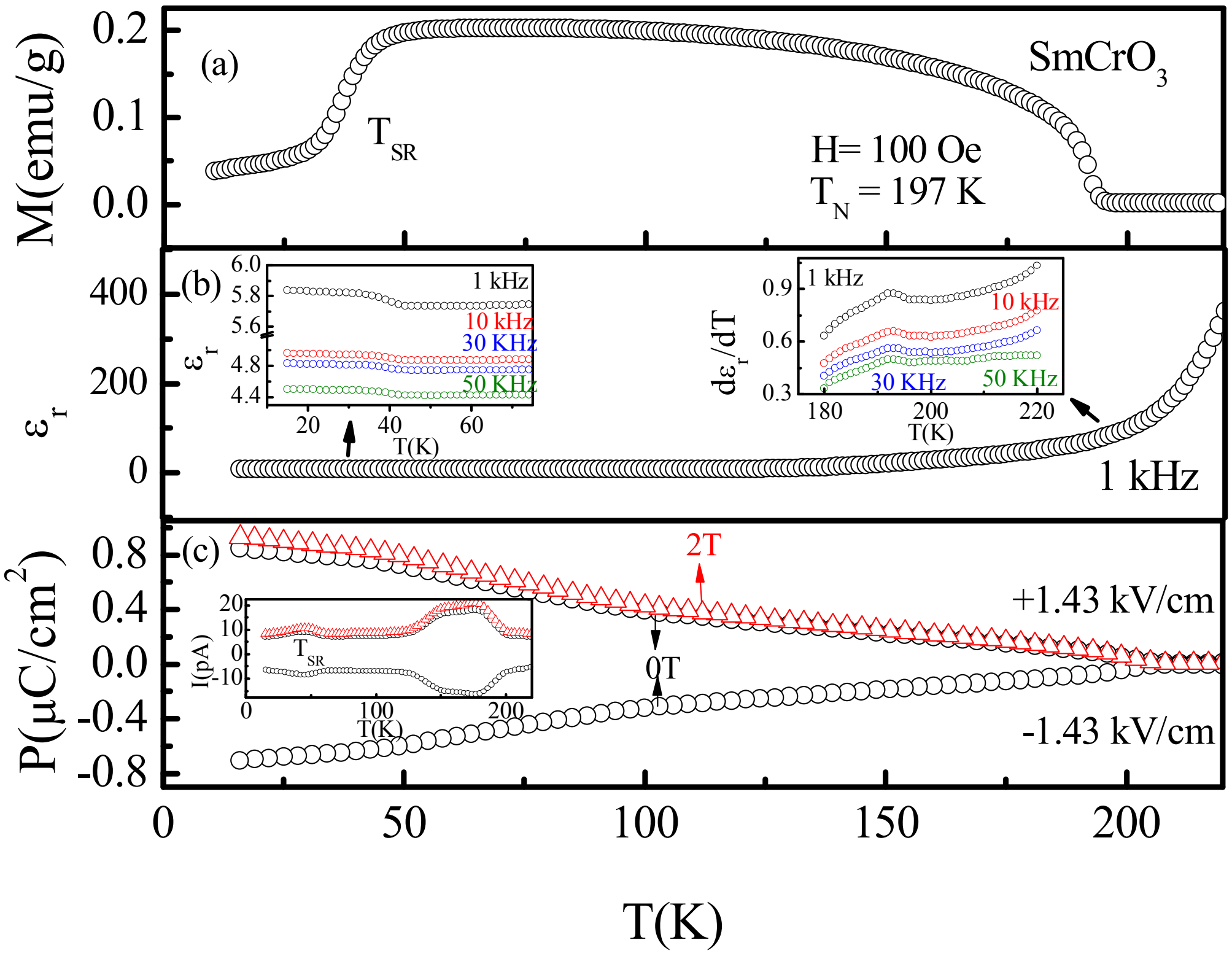

Fig. 2

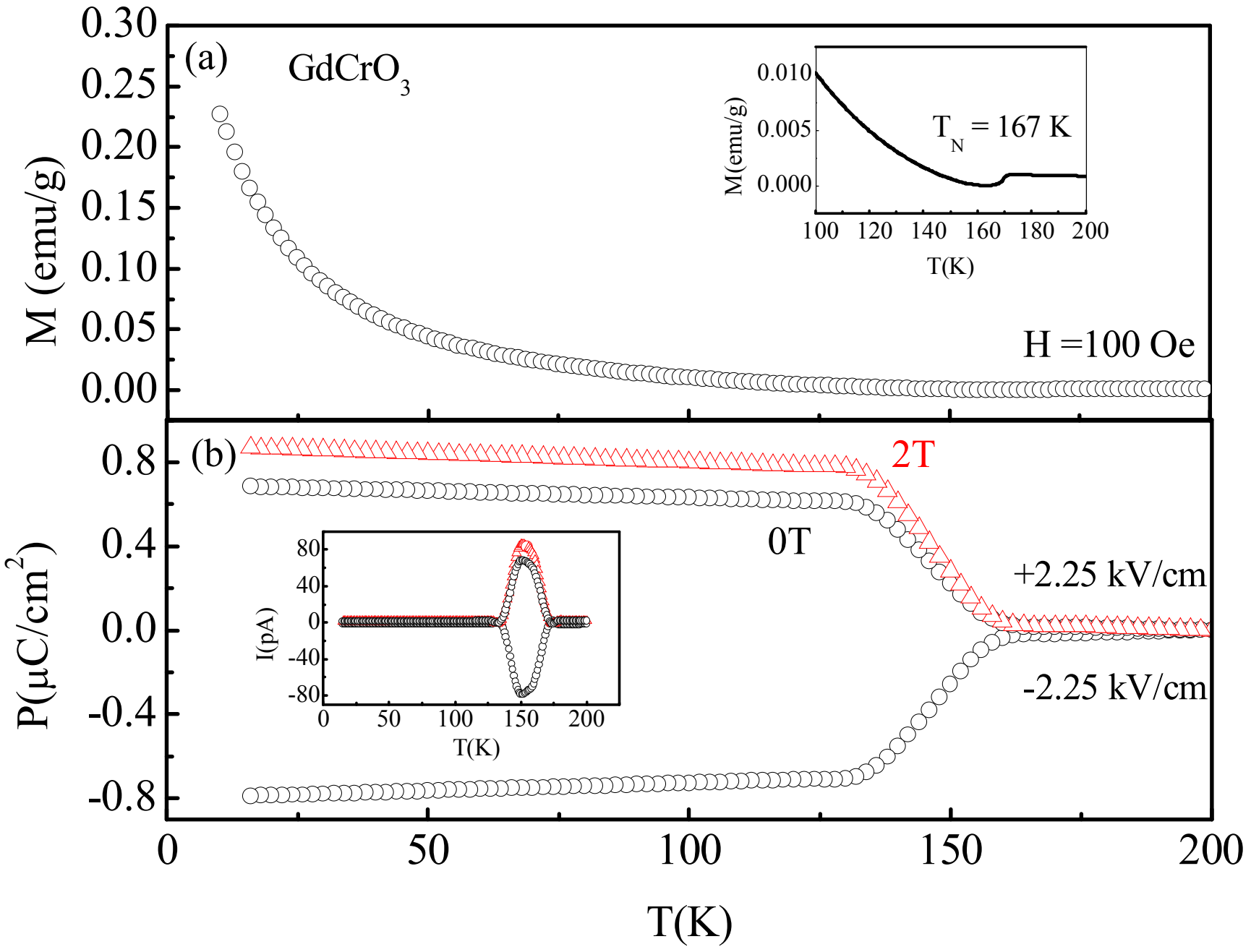

Fig. 3

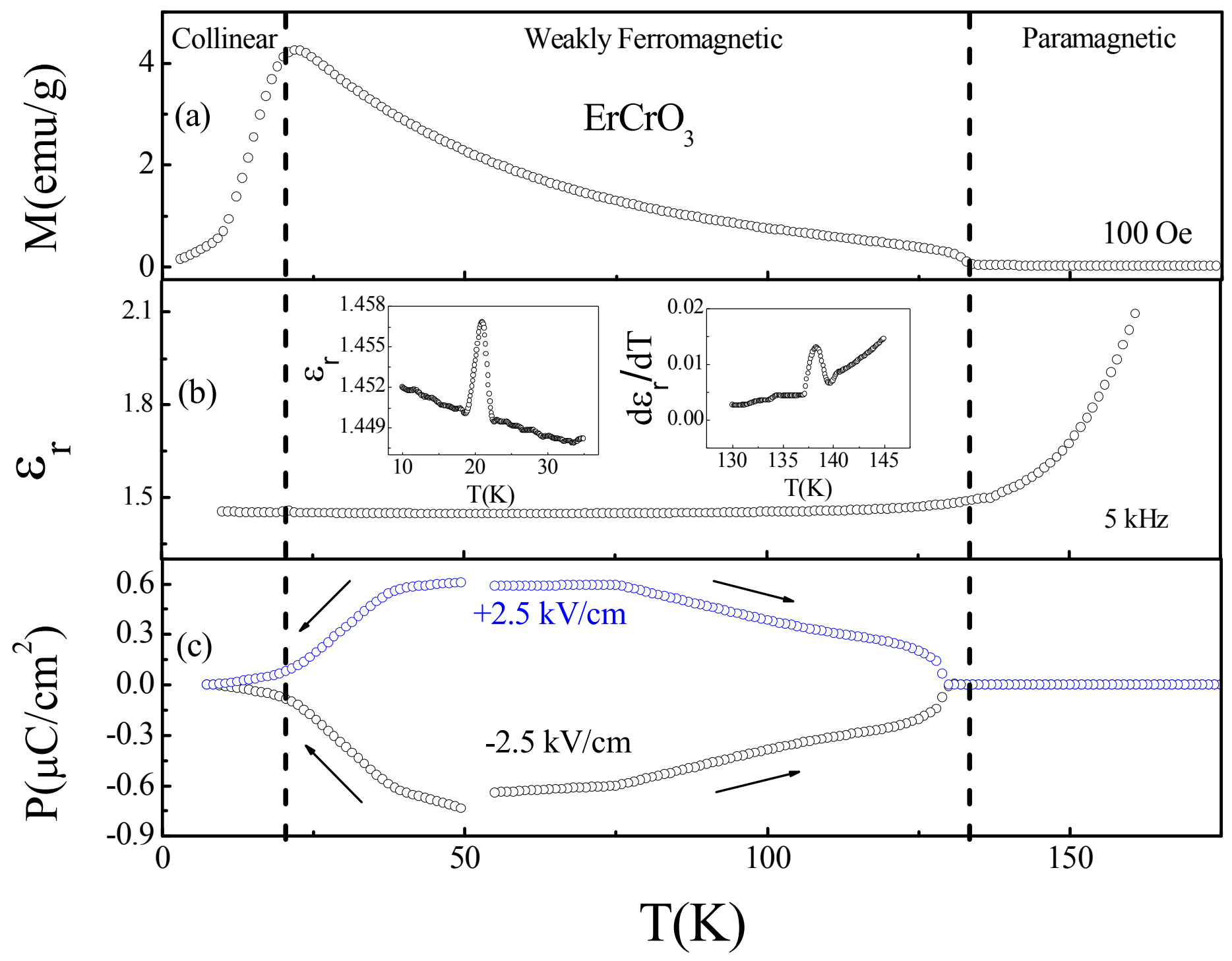

Fig. 4

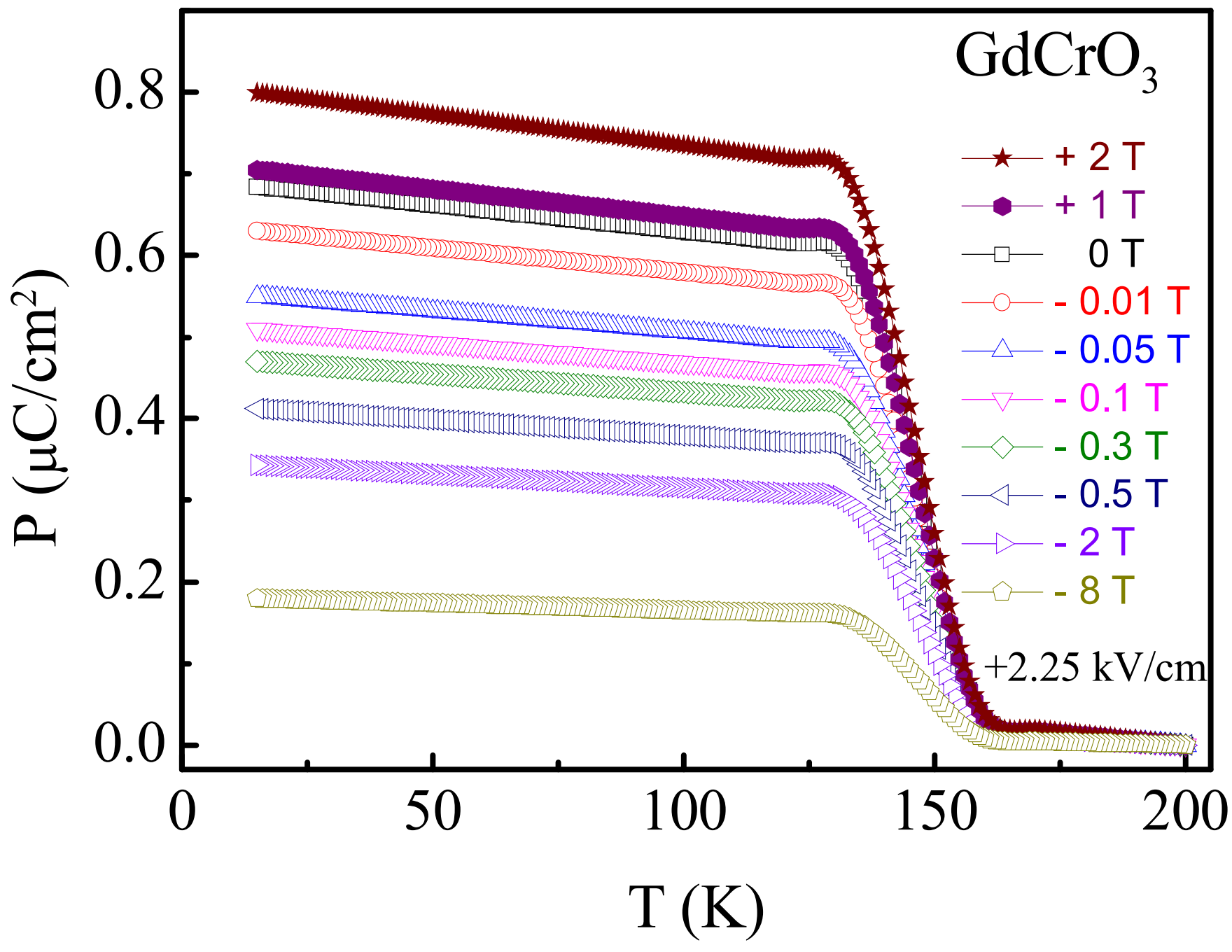

Fig. 5

# Supplemental Material

# for

# Field-induced Polar Order at the Néel Temperature of Chromium in Rare-earth Orthochromites: Interplay of Rare-earth and Cr Magnetism

B. Rajeswaran[1], D. I. Khomskii[2], A. K. Zvezdin[3], C. N. R. Rao[1] and A. Sundaresan[1*]

*[1]Chemistry and Physics of Materials Unit and International Centre for Materials Science, Jawaharlal Nehru Centre for Advanced Scientific Research, Jakkur P.O., Bangalore 560 064, India*
*[2]II. Physikalisches Institut, Universität zu Köln, ZülpicherStrasse 77, 50937 Köln, Germany*
*[3]Prokhorov General Physics Institute, Russian Academy of Sciences, ul. Vavilova 38, Moscow, 119991 Russia*

* Correspondence and requests for materials should be addressed to sundaresan@jncasr.ac.in

## Supporting Information

Positive UP – Negative Down (PUND) measurement are performed using a standard five-pulse train on the sample which enables to capture both switching (remanent + nonremanent) and non-switching (non-remanent) polarization. Pulse width, pulse delay and voltage($V_{max}$) are variable depending upon the sample to be measured. Initially, a preset pulse of $-V_{max}$ is applied on the sample while no polarization measurement was performed. Polarization measurement are started when a zero voltage pulse is applied on the sample for a fixed pulse width. After that a positive voltage ($+V_{max}$) pulse is applied which results in an increase in the polarization value. It is to be noted that this polarization is a sum total of all contributions such as that due to sample conductivity, leakage and intrinsic polarizations, stray polarizations etc. The intrinsic hysteretic contribution comes from the fact that the $V_{max}$ is switched from negative to positive values. The intrinsic polarization here can be called P*(to follow the convention). A zero pulse is again applied and the polarization measured ($P^*_r$) at this stage does not exactly go back to zero if the sample has remanence. But it is to be noted that it can also have a value due to leakage effects and conductivity. Another pulse of the same $+V_{max}$ is applied after this step and the measured polarization P^ is only due to extrinsic effects (leakage, conductivity, stray currents etc) because there is no switching of voltage polarity. After that, the voltage is set to zero and the measured polarization is assigned as $P\^{}_r$. A similar measurement routine is performed when the voltage is set to $-V_{max}$ after this measurement and we obtain corresponding –P*, $-P^*_r$, -P^ and $-P\^{}_r$. When we subtract the extrinsic terms like +P^, $+P\^{}_r$, -P^ and $-P\^{}_r$ from overall terms( intrinsic+extrinsic) like +P*, $+P^*_r$, –P* and –$P^*_r$, we obtain only the intrinsic polarization of the sample. The four difference terms (+P*-(+P^)), (+P*$_r$-(+P^$_r$)), (-P*-(-P^)) and (-P*$_r$-(-P^$_r$)) give identical values of intrinsic polarization of the sample.

In $SmCrO_3$, we performed the PUND measurement at two temperatures above the ferroelectric transition temperature and below the transition temperatures. Above the transition temperature (220 K), we see that there is no difference between the values of first peak value (+P*) and the second(+P^). Similar profile is seen in the negative direction also proving that there is no intrinsic polarization in the sample. Below the transition temperature (50 K), we observe a net remanent polarization which is consistent in all the difference terms mentioned previously verifying the intrinsic nature of the polarization.

The measurement clearly demonstrates the intrinsic nature of the polarization removing any further doubts about the spurious effects such as leakage currents and conductivity of the sample. PUND is an excellent tool at least to qualitatively verify the intrinsic polarization of the sample. While $SmCrO_3$ shows intrinsic polarization below $T_N$, $YCrO_3$ (see Figure S4) shows no net intrinsic polarization above or below the ordering temperature and the polarization observed is *entirely due to leakage* indicating the lack of ferroelectricity consistent with our conclusions. Figure S5 shows the ferroelectric polarization in $TmCrO_3$at the magnetic ordering temperature. Figure S6 shows the experimental data of orthochromites, $LuCrO_3$ and $YCrO_3$ with nonmagnetic A-site cations. The measured leakage and pyroelectric currents in $LuCrO_3$ and $YCrO_3$ are shown in the inset. Leakage current was measured by going to the lowest temperature and applying an electric field and waiting for a similar span of time as that during the measurement of pyrocurrent and then measuring the current as we warm the sample up. We note from the insets of this figure that both the pyroelectric current (filled circles) and leakage current (open triangles) show broad features with similar characteristics (The magnitude of the current is low and almost the same for both leakage and pyro contributions in contrast to magnetic $R^{3+}$ ions where the pyrocurrent are an order of magnitude greater than the leakage current) indicating that *leakage contributes entirely* to the observed current that confirms the absence of ferroelectricity. Similar features were observed for negative electric fields as well. This study clearly demonstrates that a peak in pyroelectric current does not always represent ferroelectricity.

Figure S7 shows the reversal in the polarization in $SmCrO_3$ and $GdCrO_3$ when the polarity of the applied magnetic field is switched. This elucidates the magnetoelectric coupling in these orthochromites. It can be seen from Fig. S8 that the spin reorientation in $ErCrO_3$ is suppressed upon the application of a magnetic field as evidenced from the field cooled magnetization measurements.

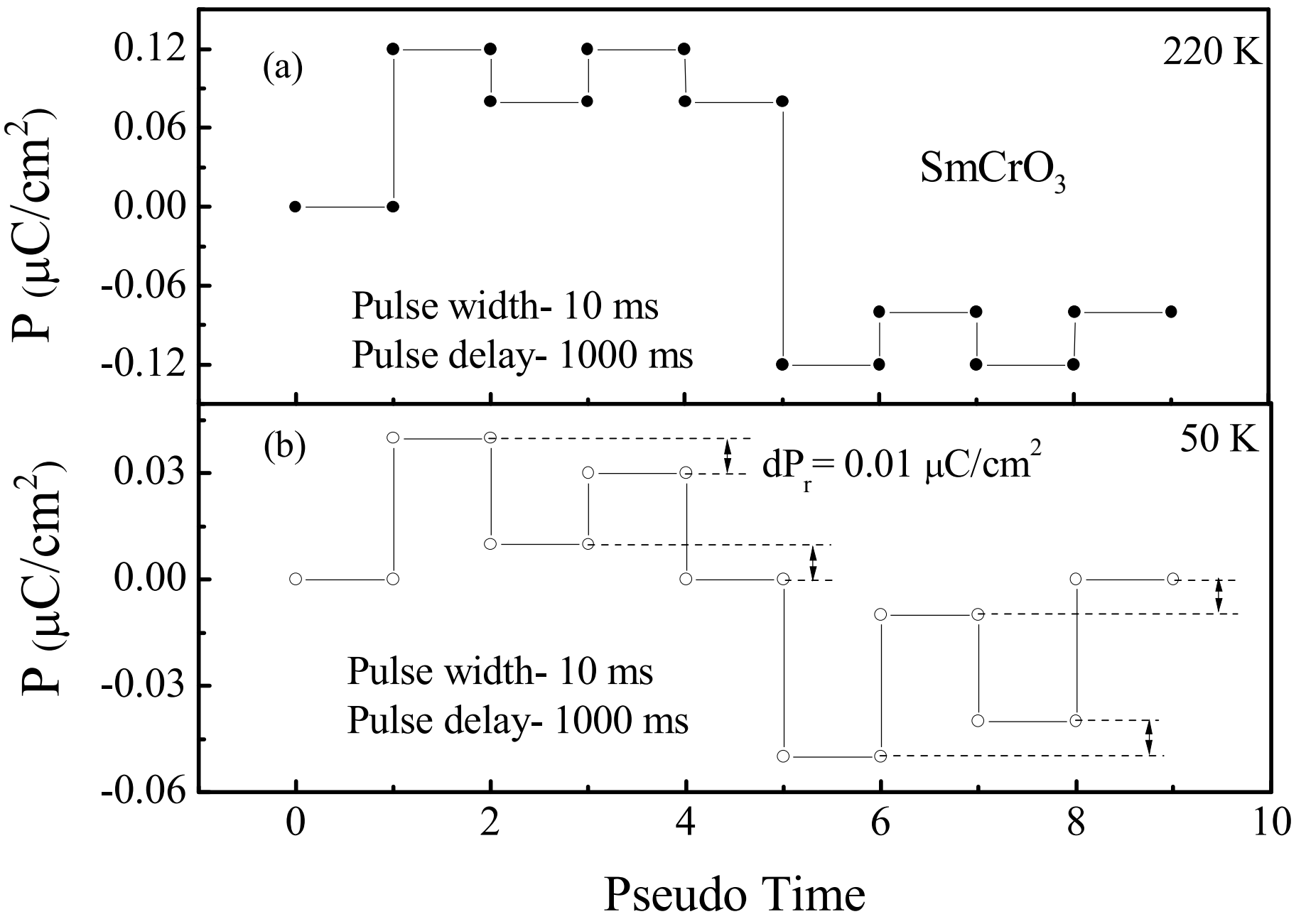

**S1** (Color Online): PUND (positive-up & negative-down) pulse results for polycrystalline $SmCrO_3$ obtained by employing the 50 V pulse of width 10 ms and the delay time of 1000ms. (a) At T = 220 K (above $T_N$), there is no remanent polarization but (b) at 50 K the presence of remanent polarization (0.01 $\mu C/cm^2$) confirms ferroelectricity. The arrows show the net intrinsic polarization in the sample.

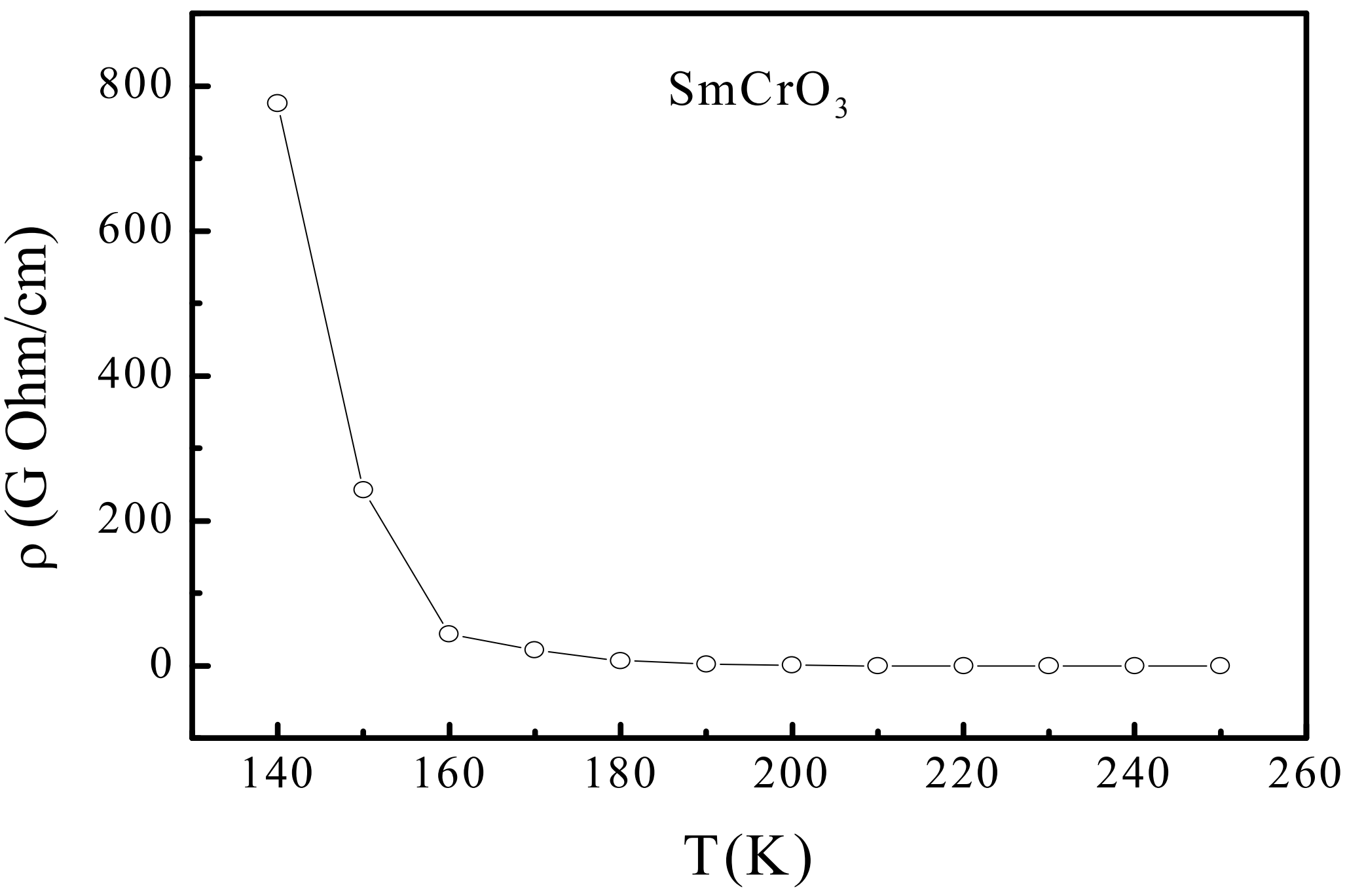

**S2** (Color Online): Resistivity of the polycrystalline $SmCrO_3$ sample as a function of temperature. It could be seen that the sample resistivity is in GOhm range at the transition temperatures. Resistivity is measured using the leakage procedure incorporated in Radiant Technologies Inc. at an applied voltage of 50 V with a soaking time of 1000 ms and a measurement time of 1000 ms.

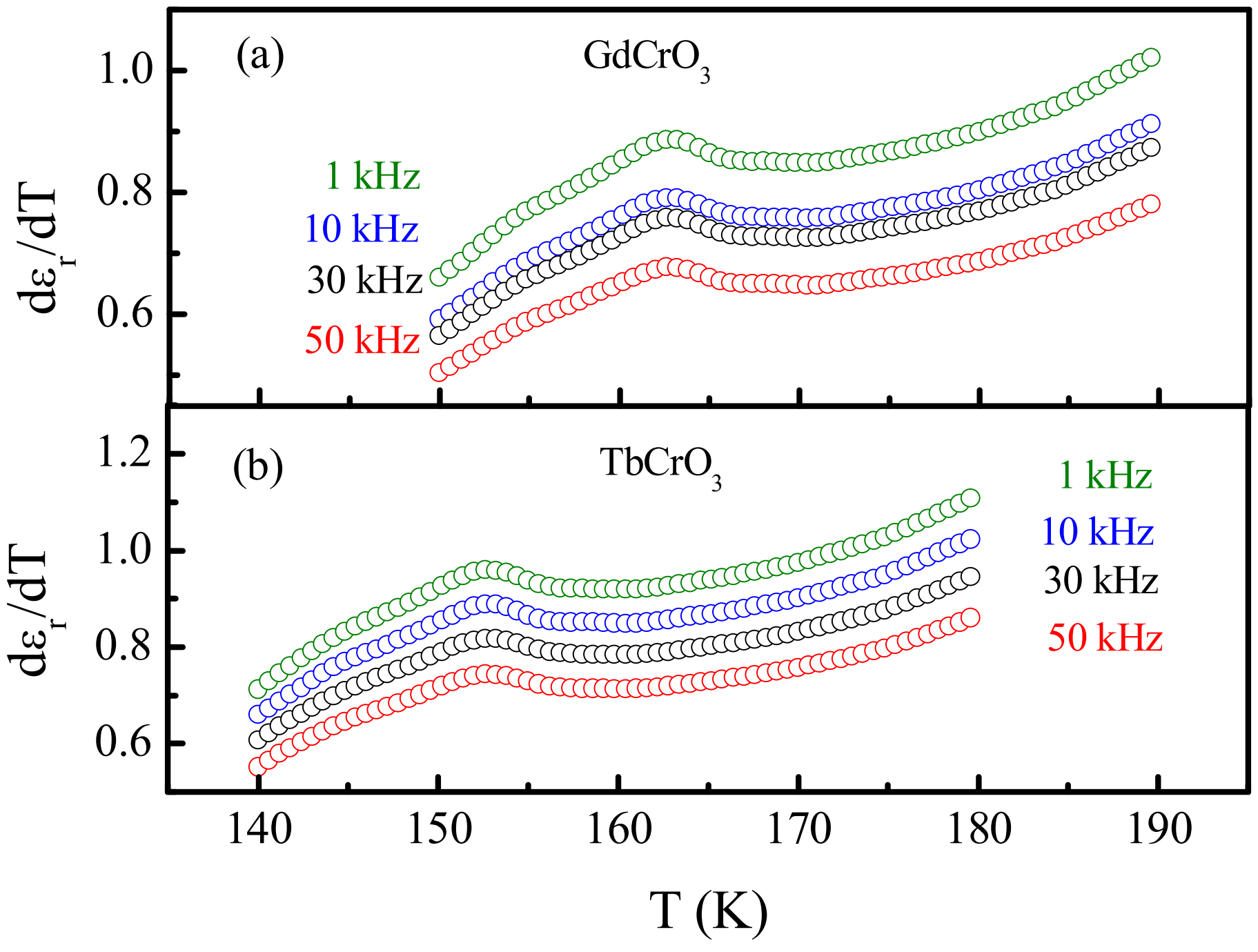

**S3** (Color online):Frequency dependence of the first derivative of dielectric constant in the vicinity of the magnetic ordering of Cr in (a) $GdCrO_3$ and (b) $TbCrO_3$. Since the temperature regime is above the onset of Maxwell-Wagner relaxation, the dielectric constant as such did not show a distinct anomaly. But it could be seen that there is a distinct anomaly in the first derivative of dielectric constant at the $T_N$ in both the cases, which demonstrates magnetodielectric effect.

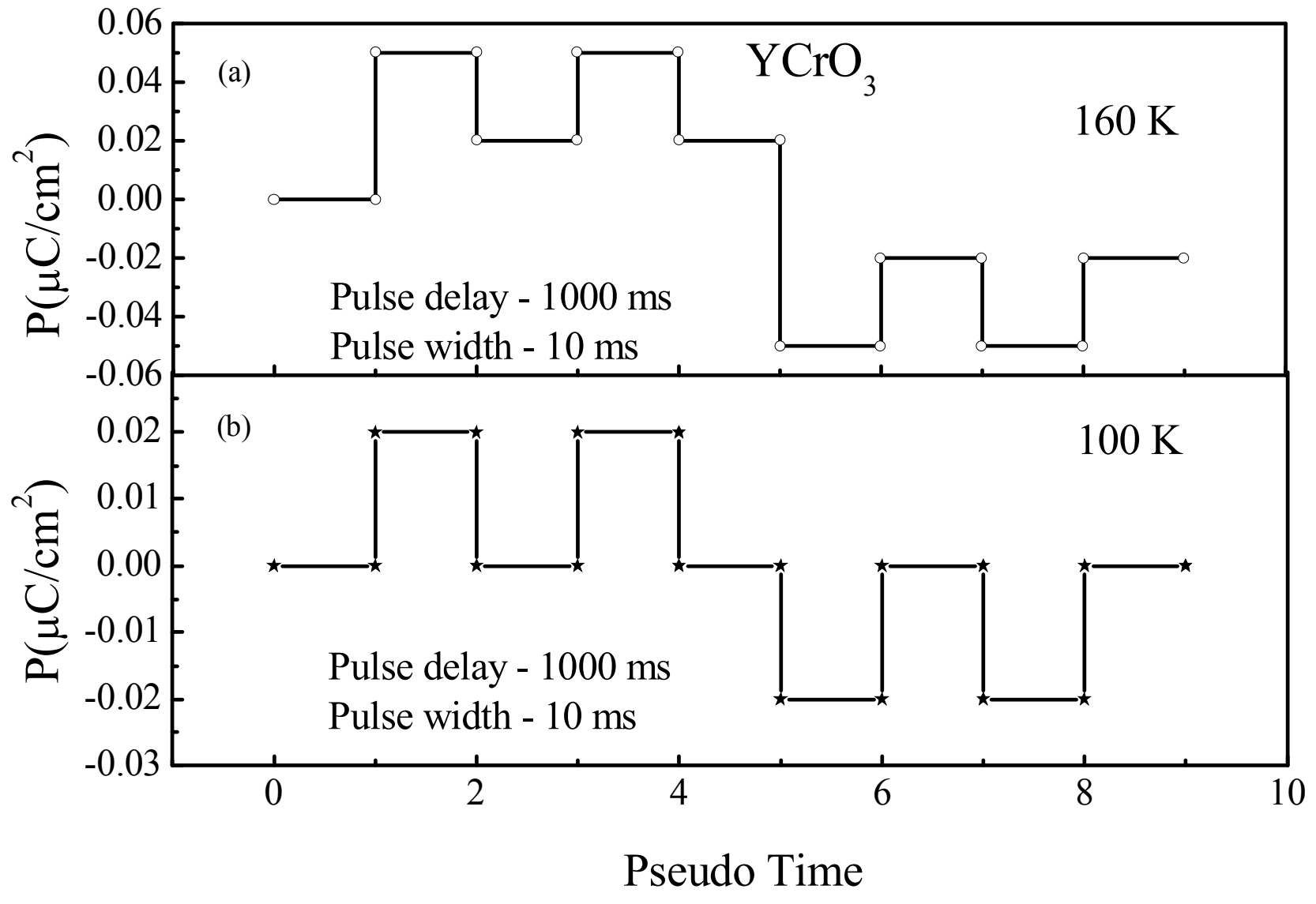

**S4** (Color Online): PUND (positive-up & negative-down) pulse result of polycrystalline $YCrO_3$ at (a) 160 K (above $T_N$) and (b) 100 K (below $T_N$). It can be seen that leakage contributes entirely and there is no remnant polarization $P_r$ over the entire temperature range demonstrating the absence of ferroelectricity.

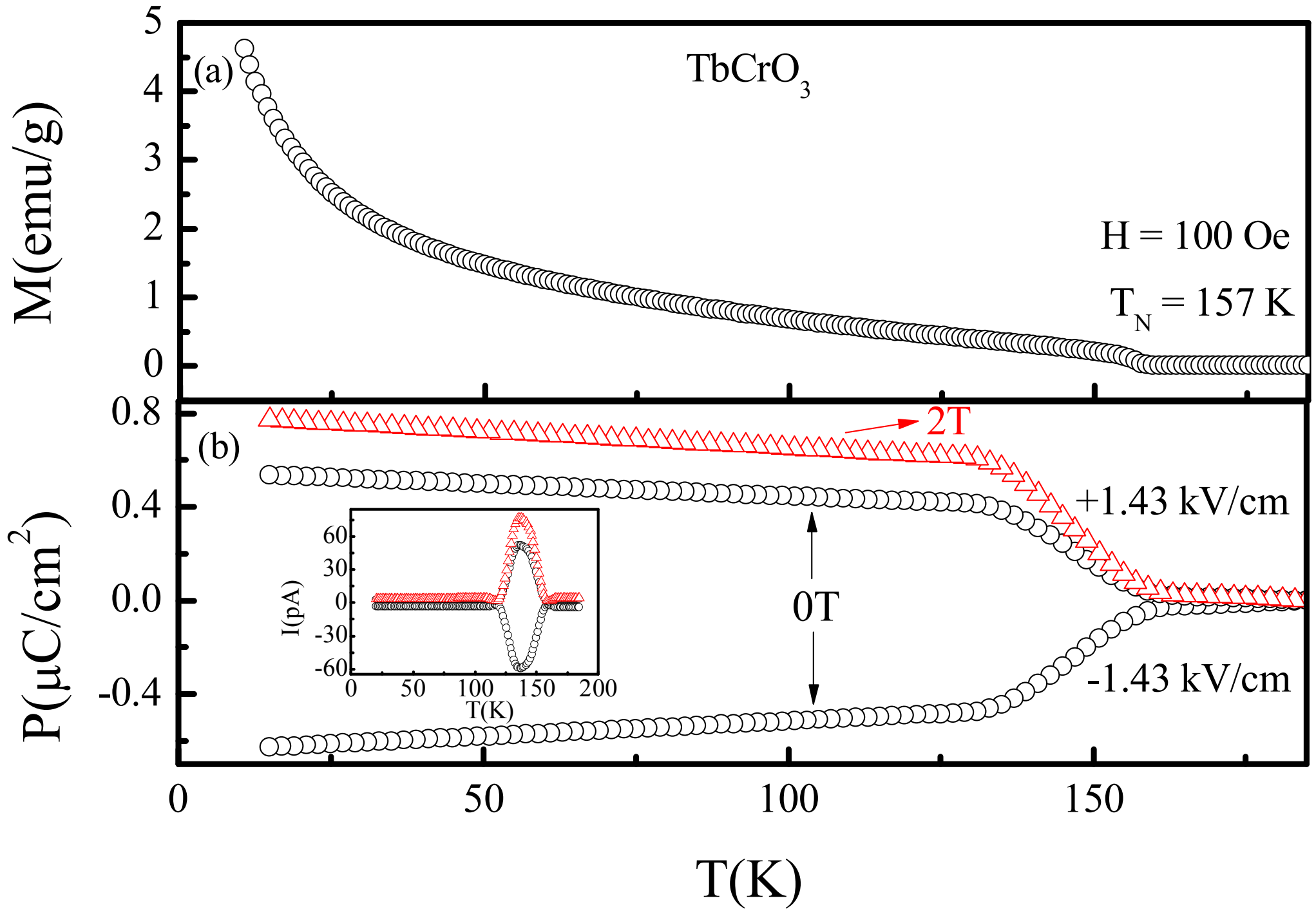

**S5**. (Color online) (a) Field-cooled magnetization of $TbCrO_3$ at 100 Oe with respect to temperature. (b) Electric polarization (corrected for leakage) as a function of temperature. Poling is done at two different fields +1.43 kV/cm and -1.43 kV/cm (black) at 0 T magnetic field and 2 T magnetic field (red). Inset in (b) shows the pyroelectric current as a function of temperature at two different poling fields with (red) and without the presence of magnetic field of 2 T (black).

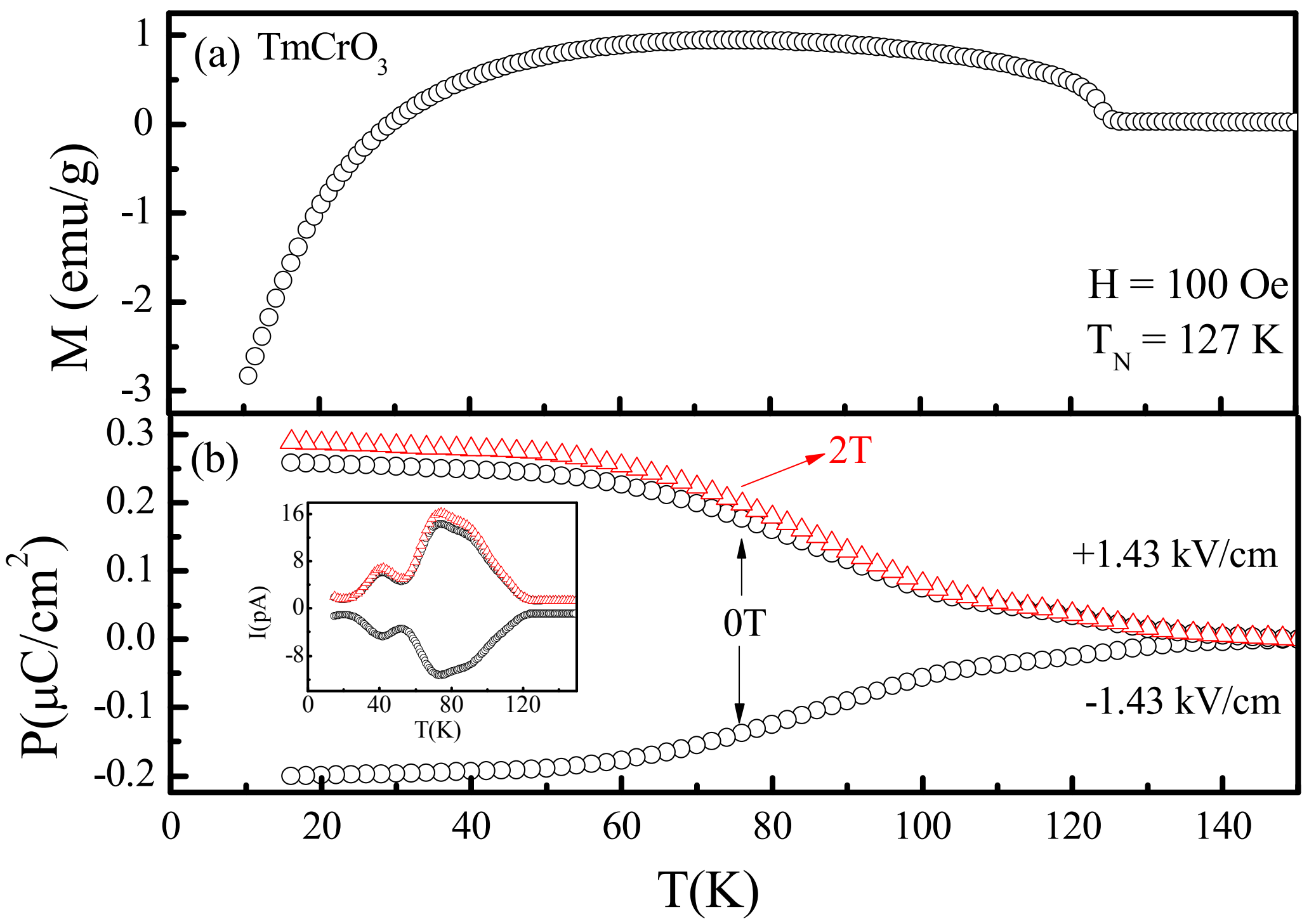

S6: (a) Field-cooled magnetization of $TmCrO_3$measured at 100 Oe showing temperature induced magnetization reversal. (b) Electric polarization (corrected for leakage) as a function of temperature. Poling is done at two different fields +1.43 kV/cm and -1.43 kV/cm (black) at 0 and 2T field (red). Inset in (b) shows the pyroelectric current as a function of temperature at two different poling fields with (red) and without the presence of magnetic field of 2T (black).

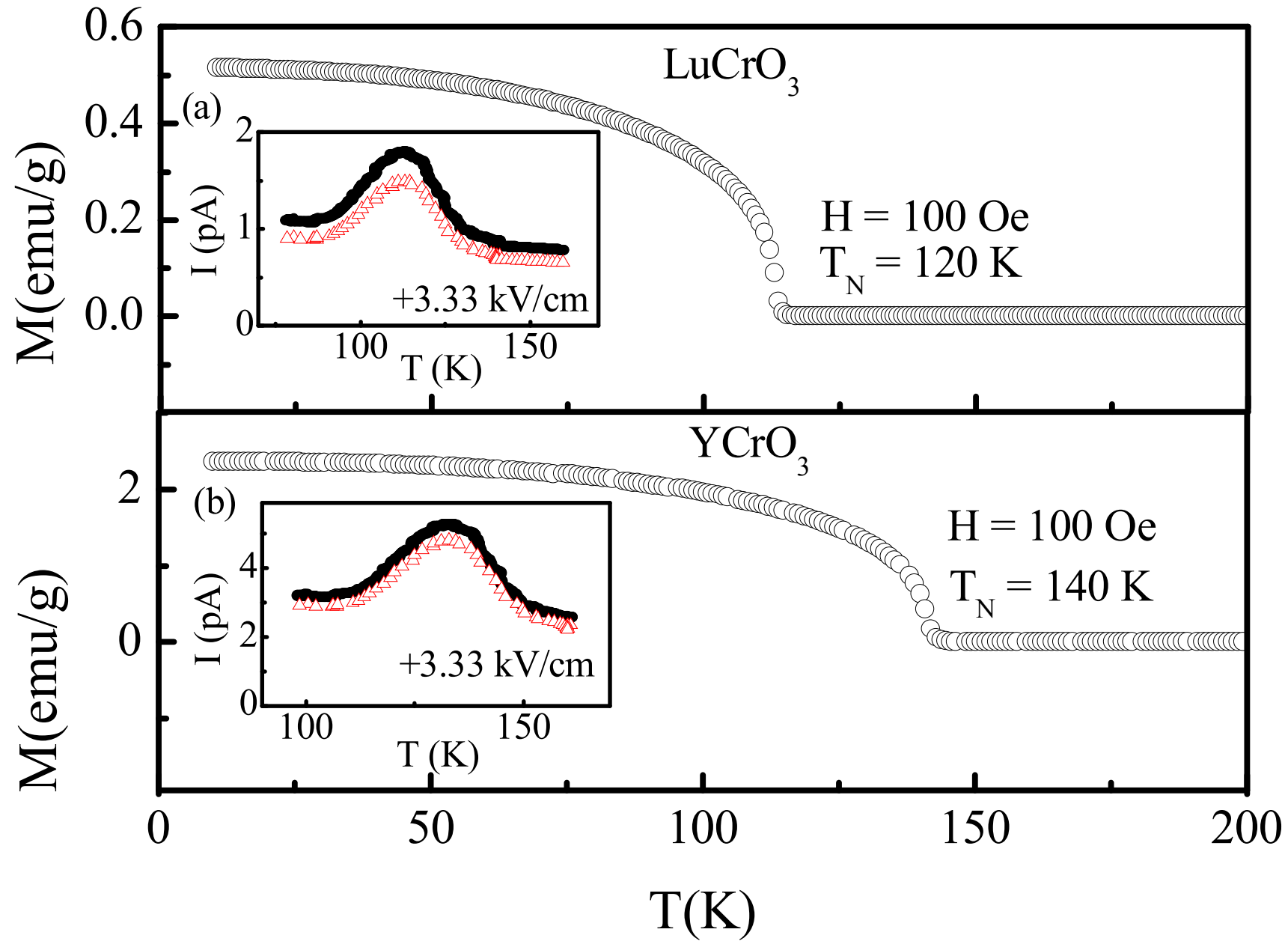

S7: Field-cooled magnetization of (a) $LuCrO_3$ and (b) $YCrO_3$at an applied field 100 Oe. Inset in both panels shows the pyroelectric current (black filled circles) with the poling field of +3.33 kV/cm and leakage current (red open triangles) as a function of temperature at similar applied DC field. It could be seen that the small peak like feature observed near $T_N$ is not due to ferroelectric polarization as the leakage contributes entirely to the observed pyroelectric current. It should also be noted that the values of the current observed is of an order of magnitude less than that for magnetic $R^{3+}$ ions.

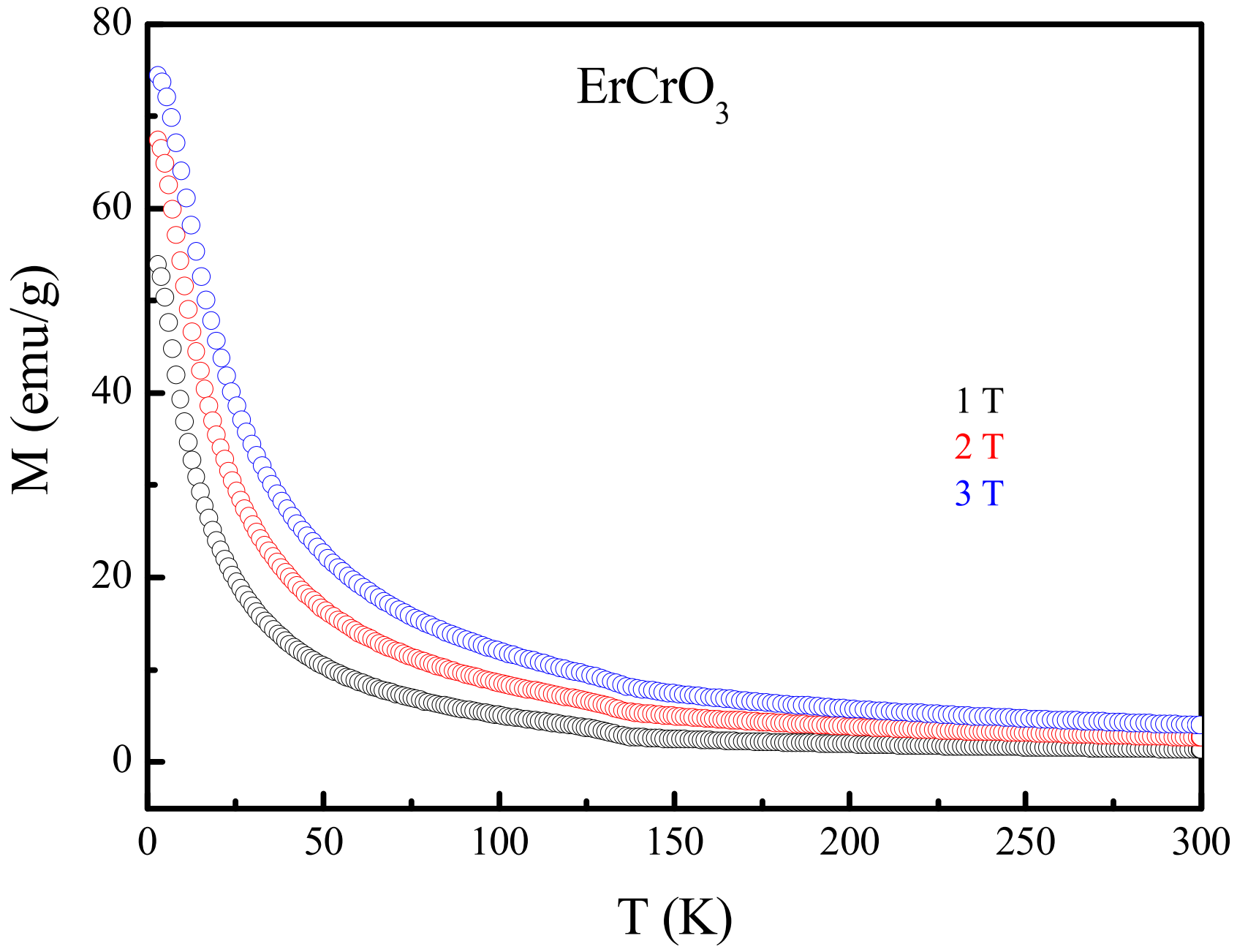

S8: Field cooled magnetization in $ErCrO_3$ demonstrating the disappearance of spin reorientation transition from the canted antiferromagnetic $\Gamma_4$ to the collinear $\Gamma_1$ phase at higher magnetic fields.